\documentclass[12pt]{article}
\usepackage{graphics}
\usepackage{graphicx}
\DeclareGraphicsExtensions{.pdf}
\usepackage{float} 
\usepackage{amsmath}
\usepackage{slashed}
\usepackage{amsfonts}   
\usepackage{amssymb}

\usepackage{color}

\begin{document}
\date{}
\title{{\bf{\Large Factorised scattering and (non)integrability for marginally deformed $ AdS_5/CFT_4 $}}}
\author{
 {\bf {\normalsize Dibakar Roychowdhury}$
$\thanks{E-mail: dibakar.roychowdhury@ph.iitr.ac.in}}\\
 {\normalsize  Department of Physics, Indian Institute of Technology Roorkee,}\\
  {\normalsize Roorkee 247667, Uttarakhand, India}
\\[0.3cm]
}

\maketitle
\begin{abstract}
We study non-SUSY CFTs that are marginal deformations of $ \mathcal{N}=2 $ SCFTs in 4d. We explore the fate of integrability and/or non-integrability in these CFTs using the techniques of factorisation of scattering amplitudes in 2d sigma models. We show possibilities for the factorisation of scattering amplitudes in the absence of flavour D6 branes which rules out non-integrability in the dual non-SUSY CFTs.
\end{abstract}
\section{Introduction and Overview}
Understanding the semiclassical quantisation of rotating superstrings in $AdS_5 \times S^5$ \cite{Berenstein:2002jq}-\cite{Metsaev:2002re} and its connection with the dual operator spectrum \cite{Gubser:2002tv}-\cite{Frolov:2002av} has been one of the major thrust areas in modern theoretical physics for last three decades. While string theory in the ``semiclassical'' limit seems to be well under control, the dual gauge theory counterpart becomes hard to deal with due to its nonperturbative nature. These questions are equally important if one goes beyond the standard Maldacena conjecture, where there is a plethora of holographic dualities starting from $AdS_7$ \cite{Apruzzi:2013yva}-\cite{Nunez:2018ags} down to $AdS_6$ \cite{DHoker:2016ujz}-\cite{Legramandi:2021uds} and $AdS_5$ \cite{Sfetsos:2010uq}-\cite{Gaiotto:2009gz}. For related works in the literature, we further refer to the reader \cite{Cremonesi:2015bld}-\cite{Macpherson:2024frt}.

The purpose of this paper is to understand the gauge/string correspondence for a particular class of type IIA superstring background, that preserves an $AdS_5$ factor and are generated through an action of $SL(3,R)$ in 11d. The 11d supergravity background preserves $ SU(2)_R \times U(1)_r $ that is dual to $ \mathcal{N}=2 $ SCFTs in 4d. In the process of dimensional reduction along $U(1)$ cycle, it mixes the R-symmetry with the 11th dimension, which breaks the SUSY completely in 10d. In other words, the $ U(1) $ of the type IIA solution is different from the $ U(1)_r $ of the parent M-theory background.  On the other hand, the $SU(2)_R$ in 10d descends from the $R$- symmetry of the parent 11d theory \cite{Macpherson:2024frt}. On the field theory counterpart, this has a consequence of adding a marginal operator (or deformation) in the parent $\mathcal{N}=2$ theory that takes to a $\mathcal{N}=0$ fixed point. 

These are one parameter ($\xi$) family of solutions, that smoothly reduce to $\mathcal{N}=2$ Gaiotto-Maldacena (GM) solution \cite{Lozano:2016kum} in the limit of the vanishing deformation ($\xi = 0$). The most interesting observation comes when considering an expansion in the region $ \sigma \sim 0 $ and $ \eta \sim P $, where $ P $ is the location of flavour D6 branes for $ \xi \neq 0 $. These geometries are known as ``single kinks'' where the kink is appearing at the location of the flavour D6 which has its role in the completion of the spacetime \cite{Lozano:2016kum}. These geometries are dual to non-SUSY CFTs where the flavour is placed at one end of the chain of color nodes. However, the above comments can be generalised for multiple stack of flavour D6 branes located at different positions ($ \eta_i $) along the holographic axis, that corresponds to multiple kinks. It turns out that the non zero D6 brane charge results in a conical defect at the poles of the internal two sphere \cite{Macpherson:2024frt}, known as ``spindles'' that had drawn renewed attention in the recent years \cite{Ferrero:2020twa}-\cite{Ferrero:2021ovq}. These orbifold singularities, that are triggered due to the marginal deformations ($ \xi \neq 0 $) of generic GM backgrounds \cite{Gaiotto:2009gz}, has opened up several new directions that are worth pursuing in the realm of gauge/string duality. 

Given the state of the art, the purpose of this paper is to understand the fate of integrability and/or non-integrability in these non-SUSY CFTs at strong coupling. Generally speaking, the standard procedure is to study various string solitons that are moving in the supergravity background and explore the associated integrability and/or non-integrability. By means of the holographic correspondence, this would allow us to conjecture about the associated integrability and/or non-integrability in the dual QFT at strong coupling. 

There has been several approaches to this, for example, the Lax pair formulation of semiclassical strings \cite{Bena:2003wd}-\cite{Kameyama:2015ufa}, analytic non-integrability of strings following Kovacic's algorithm \cite{Basu:2011fw}-\cite{Giataganas:2013dha} and so on. However, the purpose of this paper is pursue a less explored and yet an effective path namely the method of factorisation of scattering amplitudes following the lines of \cite{Wulff:2019tzh}-\cite{Giataganas:2019xdj}. We explore the warp factor close to the region $ \sigma \sim 0 $ and $ \eta \sim 0 $. This gives an alternative view on integrability and/or non-integrability of the sigma model by restricting the allowed space for $ 2 \rightarrow 2 $  scattering. We illustrate upon this by taking several examples including the $ \xi $-deformed Abelian T-dual (ATD), Sfetsos-Thompson (ST) and finally by incorporating flavour D6 branes into the picture. 

Following an appropriate tuning of the deformation parameter ($\xi$), we show possibilities for factorisation of scattering amplitude in both deformed ATD and ST solutions, which unveil an integrable structure on the stringy side of the correspondence. As a result, the dual CFT is conjectured to be integrable for some specific marginal deformations that are introduced in the parent $\mathcal{N}=2$ SCFTs. On the other hand, for deformed GM solutions with flavour D6, our analysis produces a negative outcome which shows non-integrability for non-SUSY CFTs carrying flavour degrees of freedom. 

The organisation for the rest of the paper is as follows. In Section 2, we discuss marginally deformed GM solutions and set different notations that are relevant in the subsequent analysis. In Section 3, we elaborate on factorisation of scattering amplitudes and its connection with integrability and/or non-integrability by taking several examples of deformed GM class. Finally, we draw our conclusion in Section 4.
\section{Deformed GM backgrounds}
We begin by introducing deformed GM backgrounds that are dual to non-SUSY CFTs in 4d. The NS sector of the type IIA background is given by \cite{Macpherson:2024frt}
\begin{align}
\label{e2.1}
&ds_{10}^2 = f_1 (\sigma , \eta) ds^2_{AdS_5}+f_2 (\sigma , \eta)(d \sigma^2 + d\eta^2)\nonumber\\
&+f_3 (\sigma , \eta)(d \chi^2 + \sin^2\chi d\varphi^2)+f_4 (\sigma , \eta)d\beta^2\\
&B_2 = f_5 (\sigma , \eta)\sin \chi d\chi \wedge d\varphi.
\end{align}

The individual metric functions $f_i (\sigma , \eta)$ read as
\begin{align}
\label{e2.3}
	&f_1 (\sigma , \eta)=4\sqrt{\Delta}\Big(\frac{2 \dot{V}-\ddot{V}}{V''} \Big)^{1/2}~;~f_2 (\sigma , \eta)=\frac{1}{2}f_1 (\sigma , \eta)\frac{V''}{\dot{V}}\\
	&f_3 (\sigma , \eta)=\frac{1}{2}f_1 (\sigma , \eta)\frac{V'' \dot{V}}{V''(2\dot{V}- \ddot{V})+\dot{V}'^2}~;~f_4 (\sigma , \eta)=\frac{\sigma^2}{\Delta}f_1 (\sigma , \eta)\frac{V''}{2\dot{V}-\ddot{V}}\\
	&f_5 (\sigma , \eta)=2 \Big( \frac{\dot{V}\dot{V}'}{V'' (2\dot{V}-\ddot{V})+\dot{V}'^2}-\eta\Big)-\frac{ 4\xi \dot{V}^2 V''}{V'' (2\dot{V}-\ddot{V})+\dot{V}'^2}\\
	&\Delta = \Big( 1+\frac{2 \xi \dot{V}\dot{V}'}{2\dot{V}-\ddot{V}}\Big)^2 + \frac{2 \xi^2 \sigma^2 V'' \dot{V}}{(2 \dot{V}-\ddot{V})^2}\Big(V'' (2 \dot{V}-\ddot{V})+\dot{V}'^2\Big).
	\label{e2.6}
\end{align}

Here, $V(\sigma , \eta)$ is the potential function, that characterises generic GM class of solutions \cite{Gaiotto:2009gz} and satisfies the 3d cylindrical symmetric Laplace's equation \cite{Nunez:2019gbg}
\begin{align}
	\ddot{V}+\sigma^2 V'' =0
\end{align}
where we denote, $\dot{V}=\sigma \partial_\sigma V$ and $V' = \partial_\eta V$.

Below, we illustrate upon with some simple examples of $V(\sigma , \eta)$, including the $\xi$- deformed ATD and $\xi$- deformed ST. Clearly, in the limit of the vanishing deformation ($\xi \rightarrow 0$), one smoothly recovers the $\mathcal{N}=2$ solution of \cite{Roychowdhury:2021eas}. 

The potential function for ATD reads as \cite{Lozano:2016kum}
\begin{align}
\label{e2.8}
	V_{ATD}(\sigma , \eta)= \log \sigma -\frac{\sigma^2}{2}+ \eta^2.
\end{align}

A straightforward computation reveals the deformed metric functions near $\sigma, \eta \sim 0$ as
\begin{align}
\label{e9}
&f_1 (\sigma \sim 0 , \eta \sim 0)=4+8 \xi ^2 \sigma ^2~;~f_2 (\sigma \sim 0 , \eta \sim 0)=4(1+\sigma^2)+8 \xi ^2 \sigma ^2\\
& f_3(\sigma \sim 0 , \eta \sim 0)=(1- \sigma^2) + 2 \xi^2 \sigma^2\\
&f_4(\sigma \sim 0 , \eta \sim 0)= 4 \sigma^2 ~;~ f_5(\sigma \sim 0 , \eta \sim 0)=-2 \eta -2\xi (1-2 \sigma^2)
\label{e11}
\end{align}
where we retain ourselves up to quadratic order in the expansion.

The corresponding rank function \cite{Lozano:2016kum} appears to be constant
\begin{align}
	\mathcal{R}(\eta)=1.
\end{align}
The corresponding non-SUSY CFT is a circular one containing finite number of $ SU(N_c) $ colour nodes of uniform rank that are connected through bi-fundamentals.

The ST solution is characterised by the following potential function \cite{Nunez:2018qcj}
\begin{align}
\label{e2.13}
	V_{ST}(\sigma , \eta)=\frac{\eta ^3}{3}-\frac{\eta  \sigma ^2}{2}+\eta  \log \sigma .
\end{align}

The associated metric functions, when expanded close to $\sigma, \eta \sim 0$ reveals
\begin{align}
\label{e2.14}
	&f_1 (\sigma \sim 0 , \eta \sim 0)=4(1+ \xi)-\frac{2\xi \left(3 \xi +4  \right)\sigma^2}{(1+\xi )}\\
	&f_2 (\sigma \sim 0 , \eta \sim 0)=4(1+ \xi)-\frac{2 \left(\xi ^2-2\right)\sigma^2}{(1+\xi )}\\
	&f_3 (\sigma \sim 0 ,\eta \sim 0)=4 \eta ^2(1+ \xi)\\
	\label{e2.17}
	&f_4 (\sigma \sim 0 ,\eta \sim 0)=\frac{4 \sigma ^2}{(1+\xi)} \\ 
	&f_5 (\sigma \sim 0 ,\eta \sim 0)=\mathcal{O}(\eta^3)
	\label{e18}
	\end{align}
where we retain ourselves up to quadratic order in the expansion.

The associated rank function yields a constant slope
\begin{align}
	\mathcal{R}(\eta)=\eta
\end{align}
which shows that the rank of the colour $SU(N_c)$ gauge group in the dual $\mathcal{N}=0$ CFTs increases linearly. In other words, like in the previous example, the CFTs is unbounded and hence one expects large values for various physical entities unless one sets an upper cut-off along the holographic $\eta$- axis \cite{Roychowdhury:2021eas}. This hard cut-off would act like a source namely the $O_6$- planes. The way to see this $O_6$- plane is to consider an expansion of the metric functions close to $\eta=P$ and $\sigma=0$ \cite{Lozano:2019emq}. One can make these CFTs of finite size by adding appropriate flavour nodes at the end points of the long chain of colours. This corresponds to adding flavour D6 branes in the dual supergravity description. However, as \cite{Nunez:2018qcj} reveals, adding flavour D6 spoils the integrability of the soliton, which is considered to be a generic feature of different string backgrounds in the GM class and is expected to be true in the presence of $\xi$- deformation. We confirm this claim following the approach of factorised scattering in an example with Maldacena-Nunez (MN) solution \cite{Maldacena:2000mw}.
\section{Factorised scattering}
We now elaborate on the factorisation of scattering amplitudes on the stringy side of the correspondence and its implications on the integrability and/or non-integrability of $\mathcal{N}=0$ CFT. To understand this properly, one has to rewrite the background \eqref{e2.1} as
\begin{align}
\label{e4.1}
ds^2_{10}=e^{2A}(e^{-2\psi}(dz^+ dz^- +dz^2_m)+ d\psi^2) + h_{ij}(y)dy^idy^j
\end{align}
where $ h_{ij}(y) $ is the metric of the internal space and $z$ and $\psi$ are the coordinates of $AdS_5$.

For an ``integrable'' sigma model, the $n$- particle S matrix factorises into products of $ 2 \rightarrow 2 $ scattering amplitudes and the amplitude for the process $  zz \rightarrow y y $ vanishes if the masses of $ z $ and $ y $ particles are different. Following \cite{Wulff:2019tzh}, when one expands the sigma model about a particular vacuum (namely the null cusp solution of $AdS_5$) and the 10d background is expressed as in \eqref{e4.1}, the non vanishing of $ 2 \rightarrow 2 $ amplitude is simply dictated by the warp factor associated with $ AdS_5 $ and the metric of the internal space
\begin{align}
\label{e4.2}
e^{2A}=1+2 a_{ij} y^i y^j ~;~ h_{ij}=\delta_{ij}+\mathcal{O}(y).
\end{align}
The above expansion \eqref{e4.2} ensures that the null cusp solution exists for a given supergravity background and hence all the arguments behind factorization of scattering amplitudes hold true. Clearly, these arguments do not hold for backgrounds those fail to satisfy \eqref{e4.2}.

It turns out that for an integrable sigma model, the factorised scattering produces a non zero amplitude if $ a= \frac{1}{2} $ \cite{Wulff:2019tzh}. On the other hand, the amplitude vanishes if $ a=0 $. Therefore, the factorisation of amplitudes constrains the eigenvalues of $ a_{ij} $ to be $ \lbrace 0 ,\frac{1}{2}\rbrace $ and the sigma model is conjectured to be non-integrable for all other values of $ a_{ij} $. Below, we illustrate upon the above algorithm with some simple examples of deformed GM class.

We begin with an example of $ \xi $-deformed ATD \eqref{e2.8}. Following a suitable rescaling of the 10d metric, one could schematically express \eqref{e2.1} near $ \sigma \sim 0 $ and $ \chi \sim 0 $ as\footnote{We introduce new variables, $y_1 = \chi \cos\varphi$ and $y_2=\chi \sin \varphi$ and express the metric on the two sphere as $ d\Omega_2(\chi ,\varphi)= dy^2_1 + dy^2_2$, where we absorb the pre-factor following a rescaling of coordinates ($ y_i $).}
\begin{align}
\label{e22}
\frac{1}{4}ds^2_{10}=(1+2 \xi^2 \sigma^2)ds^2_{AdS_5}+d\sigma^2+d \eta^2 + dy^2_i ~;~i=1,2
\end{align}
where we set one of the internal space directions $ \beta=0 $, in order for the expansion \eqref{e22} to be compatible with \eqref{e4.2}. One could clearly read off the coefficient $ a_{\sigma \sigma}=\xi^2 $. Clearly, for the undeformed ($ \xi =0 $) case, the factorisation $ zz \rightarrow \sigma \sigma $ exists and the amplitude vanishes. On the other hand, if we move slightly away ($|\xi| \ll 1$) from $ \mathcal{N}=2 $ fixed point, the factorisation does not take place due to different masses of the incoming and outgoing modes. However, there exist special points $ \xi_{\pm}=\pm \sqrt{\frac{1}{2}} $ in the parameter space, where the $ 2\rightarrow 2 $ amplitude becomes non-vanishing and reveals an underlying integrable structure. We conjecture that this would correspond to adding a suitable marginal coupling (in the parent $ \mathcal{N}=2 $ SCFTs), that takes to a non-SUSY fixed point preserving integrability.

While considering the $ \xi $- deformed ST, one finds an expansion near  $ \sigma , \eta \sim 0 $
\begin{align}
\label{e23}
\frac{(1+\xi)^{-1}}{4}ds^2_{10}=\Big(1-\frac{2 \xi  (3 \xi +4)\sigma ^2}{4(1+\xi)^2}\Big)ds^2_{AdS_5}+d\sigma^2+\cdots
\end{align}
which shows that the coefficient $ a_{\sigma \sigma}=-\frac{ \xi  (3 \xi +4)}{4(1+\xi)^2} $. Clearly, setting $\xi=0$, one meets the criteria of factorised scattering \cite{Wulff:2019tzh} as alluded to the above. In other words, the undeformed case naturally meets the requirements of integrability \cite{Nunez:2018qcj}. On the other hand, setting $a_{\sigma \sigma}=\frac{1}{2}$, one finds the corresponding values for the deformation parameter to be $\xi_{\pm} =\frac{1}{5} \left(-4\pm \sqrt{6}\right)$, which shows that the factorisation of the scattering $ zz \rightarrow \sigma \sigma $ exists with a non-vanishing amplitude that leads towards integrability. Combining the above pictures together with \cite{Nunez:2018qcj}, we conclude that under a suitable addition of the marginal coupling, the $\mathcal{N}=2$ theory flows to a $ \mathcal{N}=0 $ fixed point preserving integrability. However, the above statement holds true only in the absence of flavour degrees of freedom.

As emphasised above, the integrability is lost when flavour D6 are added in the bulk GM solution. We illustrate upon this taking an example of MN solution \cite{Maldacena:2000mw}
\begin{align}
\label{e24}
&V_{MN}(\sigma , \eta)=\frac{1}{2}\Big(2 \eta  \log \sigma -\sqrt{(N-\eta )^2+\sigma ^2}+\sqrt{(\eta +N)^2+\sigma ^2} \nonumber\\
&+(N-\eta ) \log \left(-\eta +\sqrt{(N-\eta )^2+\sigma ^2}+N\right)\nonumber\\
&-(\eta +N) \log \left(\eta +\sqrt{(\eta +N)^2+\sigma ^2}+N\right) \Big).
\end{align}

Given \eqref{e24}, one can compute all the metric functions $f_i(\sigma, \eta)$ and expand them close to the origin $\sigma , \eta \sim 0$, which reveals
\begin{align}
\label{e25}
&f_1 (\sigma \sim 0, \eta \sim 0)=4\sqrt{2}(1+\xi)N-\frac{2\sqrt{2}}{N}(1+\xi)\eta^2+\frac{\sqrt{2} (3+2 \xi )\sigma^2}{(1+\xi ) N}\\
&f_2 (\sigma \sim 0, \eta \sim 0)=\frac{2\sqrt{2}}{N}(1+\xi)~;~f_3 (\sigma \sim 0, \eta \sim 0)=\frac{2\sqrt{2}}{N}(1+\xi)\eta^2\\
&f_4 (\sigma \sim 0, \eta \sim 0)=\frac{2 \sqrt{2}\sigma^2}{(1+\xi )N}
\label{e27}
\end{align}
where we retain ourselves up to quadratic order in the expansion.

Using \eqref{e25}-\eqref{e27}, we find
\begin{align}
\frac{(1+\xi)^{-1}}{4\sqrt{2}N}ds^2_{10}=\Big( 1-\frac{\eta^2}{2N^2}+\frac{(3+ 2\xi)\sigma^2}{4N^2 (1+\xi)^2} \Big)ds^2_{AdS_5}+\frac{1}{2N^2}(d\sigma^2+d\eta^2)+\cdots
\end{align}
which after a rescaling of the internal coordinates $ \sigma \rightarrow \frac{\sigma}{\sqrt{2}N} $ and $ \eta \rightarrow \frac{\eta}{\sqrt{2}N} $, yields
\begin{align}
\frac{(1+\xi)^{-1}}{4\sqrt{2}N}ds^2_{10}=\Big( 1-\eta^2+\frac{(3+ 2\xi)\sigma^2}{2 (1+\xi)^2} \Big)ds^2_{AdS_5}+d\sigma^2+d\eta^2+\cdots.
\end{align}

We have the following coefficients $ a_{\eta \eta}=-\frac{1}{2} $ and $ a_{\sigma \sigma}= \frac{(3+ 2\xi)}{4(1+\xi)^2} $. In the undeformed ($ \xi =0 $) case, they simply reduce to $ a_{\eta \eta}=-\frac{1}{2} $ and $ a_{\sigma \sigma}= \frac{3}{4} $, which forbids the factorisation of scattering amplitudes for obvious reasons. One of the coefficients ($ a_{\eta \eta} $) turns out to be negative and the other does not meet the requirements of \cite{Wulff:2019tzh}. In other words, the presence of flavour D6 leads towards non-integrability of the $ \mathcal{N}=2 $ SCFT \cite{Nunez:2018qcj}. 

In the presence of $ \xi $-deformation, $ a_{\eta \eta} $ is still the same and therefore the factorisation $ zz \rightarrow \eta \eta $ does not exist. On the other hand, $ a_{\sigma \sigma} $ can be tuned to $ \frac{1}{2} $ for $ \xi_{\pm} =\frac{1}{2} \left(-1\pm \sqrt{3}\right)$ which does lead to a factorisation of the type $ zz \rightarrow \sigma \sigma $, with some non vanishing amplitude. Therefore, to summarise, not all possible $ 2 \rightarrow 2 $ amplitudes get factorised which is a primary requirement of an integrable theory and hence the dual CFT is non-integrable.
\section{Conclusions and Outlook}
Before we conclude, some general remarks are in order. First of all, it must be emphasised that the algorithm of \cite{Wulff:2019tzh} eventually offers a method of proving non-integrability in sigma models on various string backgrounds. This idea is somewhat similar in spirit to that of the Liouvillian non-integrability \cite{Basu:2011fw}-\cite{Giataganas:2013dha}, that has been widely explored in the recent years. As our analysis reveals, the method of factorised scattering works only in the domain of small coordinates, namely in an expansion near $ \sigma , \eta \sim 0 $.  In other words, moving away from the origin would naturally spoil the integrability of the sigma model.

We further show that for the $ \xi $- deformed ATD and ST backgrounds, the algorithm \cite{Wulff:2019tzh} rules out integrability except for some finly tuned values of the deformation parameter ($ \xi_{\pm} $). However, unlike the standard Lax pair formulation, this does not prove integrability. Instead, it simply points towards the fact that those special points ($ \xi_{\pm} $) do not exhibit non-integrability in the sense of factorisation of scattering amplitudes. Finally, once the flavour D6 branes are added into the picture, the integrability is lost which is compatible with the previous analysis of \cite{Nunez:2018qcj}. We illustrate upon this with an example of MN solution, which can be trivially generalised for other flavour backgrounds in the deformed GM class.
\subsection*{Acknowledgments}
The author is indebted to the authorities of IIT Roorkee for their unconditional support towards researches in basic sciences. The author would also like to acknowledge The Royal Society, UK for financial assistance. Finally, the author acknowledges the Mathematical Research Impact Centric Support (MATRICS) grant (MTR/2023/000005) received from ANRF, India.

\end{document}